\documentclass[twoside,10pt,letterpaper]{article}
\usepackage{array}
\usepackage[dvips]{graphicx}
\usepackage{cite}
\usepackage{amssymb}
\usepackage{amsmath}
\usepackage{pstricks}
\usepackage{pst-node}
\begin{document}
%
%\volnumpagesyear{0}{0}{000--000}{2001}
%\dates{received date}{revised date}{accepted date}

\title{THE ASTUMIAN'S PARADOX}
\author{Edward W. Piotrowski\\\scriptsize Institute of Theoretical Physics,
University of Bia\l ystok,\\\scriptsize Lipowa 41, Pl 15424 Bia\l
ystok, Poland\\\scriptsize e-mail:
ep@alpha.uwb.edu.pl\vspace{1ex}\\
 Jan S\l adkowski\\\scriptsize Institute of Physics, University of Silesia, \\\scriptsize Uniwersytecka
4, Pl 40007 Katowice, Poland \\\scriptsize e-mail:
sladk@us.edu.pl}
% Add author/affliation/mailing sets as required
\date{}

\maketitle

\markboth{The Astumian's Paradox}{Piotrowski and S\l adkowski}

\pagestyle{myheadings}
% Comment this out to remove the running heads

%\keywords{Parrondo's paradox; random games; martingales; random walk; random transport.}
% Keywords have to before the abstract I'm afraid.

\begin{abstract}
We discuss some aspects of Astumian suggestions
that combination of biased games (Parrondo's paradox) can explain
performance of molecular motors. Unfortunately the model is flawed
by explicit asymmetry overlooked by the author. In addition, we
show that taking into consideration stakes allows to remove the
paradoxical behaviour of  Parrondo games.
\end{abstract}
 \vspace{5mm}
\begin{center}
{\small Motto:}
{\em Nothing gets  out of nothing.}\/\cite{janosch}
\end{center}

\noindent The celebrated Parrondo's paradox \cite{par1,par2}
consisting in asymmetrical combination of doomed games  so that
the resulting new game is not biased or there even is a winning
strategy caused much excitement and, unfortunately,
misunderstanding.
 R.~D.~Astumian \cite{ast} considers in a recent
article in {\em Scientific American}\/  a game based on the presented
below diagram. The game consist in jumping between five different
states $1,\ldots,5$:
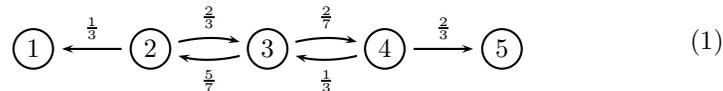
\begin{equation}
\psmatrix[mnode=circle,colsep=1]
1&2&3&4&5
\endpsmatrix
\everypsbox{\scriptstyle}
\psset{shortput=nab,nodesep=3pt,arrows=->,labelsep=3pt}
\ncline{1,2}{1,1}_{\frac{1}{3}}
\ncarc[arcangle=14]{1,3}{1,2}^{\frac{5}{7}}
\ncarc[arcangle=14]{1,2}{1,3}^{\frac{2}{3}}
\ncarc[arcangle=14]{1,4}{1,3}^{\frac{1}{3}}
\ncarc[arcangle=14]{1,3}{1,4}^{\frac{2}{7}}
\ncline{1,4}{1,5}^{\frac{2}{3}}
\vspace{.5ex}
\label{graph}
\end{equation}
where the numbers written above or below the arrows are the
probabilities of transitions between neighboring states (only such
transitions are allowed). The player wins if she (or he) winds up
in the state $5$ and loses if she  reach the state $1$. If the
player starts from the state $3$ the probability of losing is
equal to $\tfrac{5}{9}$ and winning to $\tfrac{4}{9}$. This is
because in games of this kind the proportion of probabilities of
defeat and success is given by the proportion of  products of the
appropriate transition probabilities
\begin{equation}
\frac{p_3(1)}{p_3(5)}=\frac{p(3\rightarrow2)\,p(2\rightarrow1)}{
p(3\rightarrow4)\,p(4\rightarrow5)}
\label{aba}
\end{equation}
(though it is not true that
$p_3(1)=p(3\rightarrow2)\,p(2\rightarrow1)$ nor
$p_3(5)=p(3\rightarrow4)\,p(4\rightarrow5)$). The given above
formula has a clear gambling interpretation. Consider the function
$M(n)$ (represented in graphic form in Eq\mbox{.} $(\ref{graph})$):
\begin{table}[htbp]
\caption{The  martingale corresponding to the game presented in
Eq.~$(1)$. The states $1$ and $5$ are absorbing ones.}
\centering\footnotesize
\vspace{1ex}
\begin{tabular}{|r|c|c|c|c|c|}
\hline $n=$&1&2&3&4&5\\\hline
$M(n)=$\vphantom{$\int_{j_{j_j}}^{T^T}$}&$\tfrac{-1}{p(3\rightarrow2)p(2\rightarrow1)}$&
$\tfrac{-1}{p(3\rightarrow2)}$& $0$&
$\tfrac{1}{p(3\rightarrow4)}$&
$\tfrac{1}{p(3\rightarrow4)p(4\rightarrow5)}$\\\hline
\end{tabular}
\end{table}

\noindent $M(n)$ is a martingale \cite{fel} that gives capital in a fair
game, that is in such stochastic process   that at any moment $t$:
\begin{equation}
E(M(n_{t+1})|M(n_{0}),...,M(n_{t}))=E(M(n_{t+1})|M(n_{t}))=M(n_t)\,
\end{equation}
(a finite Markov chain). The states $1$ and $5$ are absorbing,
therefore
\begin{equation}
0=M(3)=E(M(n_\infty)|n_0\negthinspace=\negthinspace3)=p_3(1)M(1)+p_3(5)M(5)\,
\end{equation}
and $(\ref{aba})$ follows.  The same numerical  result might be
obtained in  the game
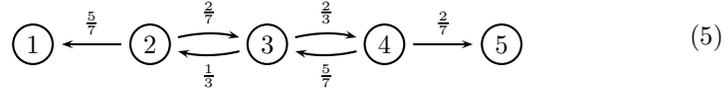
\begin{equation}
\psmatrix[mnode=circle,colsep=1]
1&2&3&4&5
\endpsmatrix
\everypsbox{\scriptstyle}
\psset{shortput=nab,nodesep=3pt,arrows=->,labelsep=3pt}
\ncline{1,2}{1,1}_{\frac{5}{7}}
\ncarc[arcangle=14]{1,3}{1,2}^{\frac{1}{3}}
\ncarc[arcangle=14]{1,2}{1,3}^{\frac{2}{7}}
\ncarc[arcangle=14]{1,4}{1,3}^{\frac{5}{7}}
\ncarc[arcangle=14]{1,3}{1,4}^{\frac{2}{3}}
\ncline{1,4}{1,5}^{\frac{2}{7}}
\end{equation}
with different transition probabilities. Astumian suggests that in
a modified game with the transition probabilities being arithmetic
mean of the probabilities of both above described games that is
the game given by
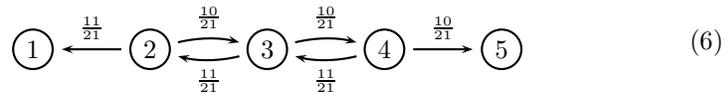
\begin{equation}
\psmatrix[mnode=circle,colsep=1]
1&2&3&4&5
\endpsmatrix
\everypsbox{\scriptstyle}
\psset{shortput=nab,nodesep=3pt,arrows=->,labelsep=3pt}
\ncline{1,2}{1,1}_{\frac{11}{21}}
\ncarc[arcangle=14]{1,3}{1,2}^{\frac{11}{21}}
\ncarc[arcangle=14]{1,2}{1,3}^{\frac{10}{21}}
\ncarc[arcangle=14]{1,4}{1,3}^{\frac{11}{21}}
\ncarc[arcangle=14]{1,3}{1,4}^{\frac{10}{21}}
\ncline{1,4}{1,5}^{\frac{10}{21}}
\end{equation}
the probability of success is greater than the probability of
defeat. He calls it the player paradox. Even a superficial
analysis of the above diagram suggests that this is wrong: the
transitions from the states $2,3,4$ to the left are more probable
than to the right. It seems that an elementary error in counting
probabilities resulted in drawing wrong conclusion by prof\mbox{.}
Astumian. There is nothing paradoxical if asymmetry put in by
hand. Roughly speaking, the Parrondo's paradox consist in the fact
that sometimes it is better to play $n$ times $game1$ followed by
$game2$ rather than $n$ times $game1$ and then $n$ times $game2$.
Below we give an example of a pair of stochastic processes biased
towards the left absorbing state and their equally weighted mixed
process biased towards the right absorbing state (Parondo's
paradox). Consider the following three graphs. 
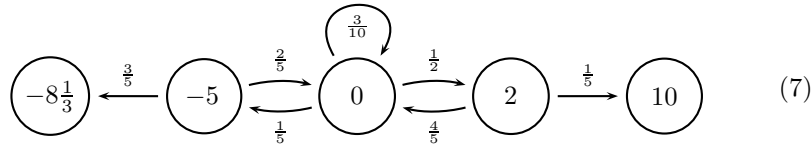
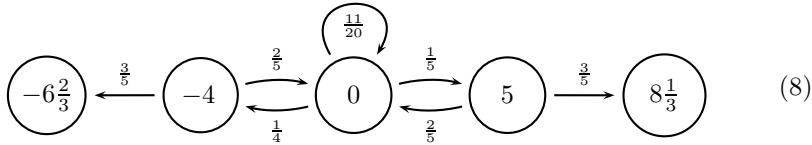
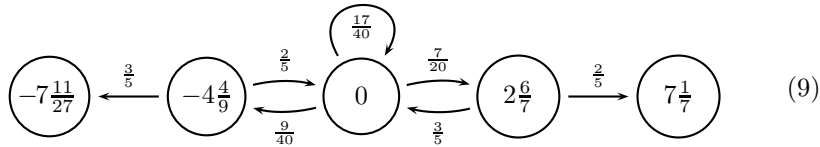
\vspace{3ex}
\begin{equation}
\psmatrix[mnode=circle,colsep=1]
-8\tfrac{1}{3}&\;-5\;\,\,&\phantom{-}0\phantom{-}&\phantom{-}2\phantom{-}&
\phantom{a}10\phantom{a}
\endpsmatrix
\everypsbox{\scriptstyle}
\psset{shortput=nab,nodesep=3pt,arrows=->,labelsep=3pt}
\ncline{1,2}{1,1}_{\frac{3}{5}}
\ncarc[arcangle=120,ncurv=3.3]{1,3}{1,3}\Bput{\frac{3}{10}}
\ncarc[arcangle=14]{1,3}{1,2}^{\frac{1}{5}}
\ncarc[arcangle=14]{1,2}{1,3}^{\frac{2}{5}}
\ncarc[arcangle=14]{1,4}{1,3}^{\frac{4}{5}}
\ncarc[arcangle=14]{1,3}{1,4}^{\frac{1}{2}}
\ncline{1,4}{1,5}^{\frac{1}{5}}\vspace{5ex}
\end{equation}
%%%%%%%%%%%%%%%%%%%%%%%%%%%%%%%%%%%%%%%%%%%%%%%%%%%%%%%%%%%%%%%%%%%%
\begin{equation}
\psmatrix[mnode=circle,colsep=1]
-6\tfrac{2}{3}&\;-4\;\,\,&\phantom{-}0\phantom{-}&\phantom{-}5\phantom{-}&
\phantom{o}8\tfrac{1}{3}\phantom{o}
\endpsmatrix
\everypsbox{\scriptstyle}
\psset{shortput=nab,nodesep=3pt,arrows=->,labelsep=3pt}
\ncline{1,2}{1,1}_{\frac{3}{5}}
\ncarc[arcangle=120,ncurv=3.3]{1,3}{1,3}\Bput{\frac{11}{20}}
\ncarc[arcangle=14]{1,3}{1,2}^{\frac{1}{4}}
\ncarc[arcangle=14]{1,2}{1,3}^{\frac{2}{5}}
\ncarc[arcangle=14]{1,4}{1,3}^{\frac{2}{5}}
\ncarc[arcangle=14]{1,3}{1,4}^{\frac{1}{5}}
\ncline{1,4}{1,5}^{\frac{3}{5}}\vspace{5ex}
\end{equation}
%%%%%%%%%%%%%%%%%%%%%%%%%%%%%%%%%%%%%%%%%%%%%%%%%%%%%%%%%%%%%%%%%%%%%%%
\begin{equation}
\psmatrix[mnode=circle,colsep=1]
\phantom{.}\hspace{-.7em}-\negthinspace\negthinspace7\tfrac{11}{27}&-4\tfrac{4}{9}&\phantom{-}0\phantom{-}&
\phantom{o}2\tfrac{6}{7}\phantom{o}&
\phantom{o}7\tfrac{1}{7}\phantom{o}
\endpsmatrix
\everypsbox{\scriptstyle}
\psset{shortput=nab,nodesep=3pt,arrows=->,labelsep=3pt}
\ncline{1,2}{1,1}_{\frac{3}{5}}
\ncarc[arcangle=120,ncurv=3.3]{1,3}{1,3}\Bput{\frac{17}{40}}
\ncarc[arcangle=14]{1,3}{1,2}^{\frac{9}{40}}
\ncarc[arcangle=14]{1,2}{1,3}^{\frac{2}{5}}
\ncarc[arcangle=14]{1,4}{1,3}^{\frac{3}{5}}
\ncarc[arcangle=14]{1,3}{1,4}^{\frac{7}{20}}
\ncline{1,4}{1,5}^{\frac{2}{5}}
\end{equation}
The numbers given at the vertices represent values of the
martingales and not their arguments.  We can identify
corresponding vertices of the mixed process if we appropriately
change the stakes (in general the stakes are different). The
formalism of martingales, besides offering a fast method of
finding probabilities of reaching the asymptotic states removes a
lot of the mysteries of the paradoxical behavior. In the mixed
process the bias can be compensated by modification of stakes. So
an increasing probability of winning not necessarily involves
increasing (expected) profits.

The authors of this letter suggest the readers to calculate the
probabilities of success in games of this kind. An exemplary
listing of a mini-program written in the language  {\em
Mathematica 5.0}\/ together with results they obtained can be
found in Appendix (cf\mbox{.} \cite{ht,wagon}).

Note that the paradox is present also in quantum games \cite{abb,ep}.

%\appendix % Reset the environments to Appendix style

\section*{Appendix}
 The following numerical calculations in {\em Mathematica 5.0}\/ show the
asymptotic behavior of the Astumian's mixed game.
\[
\begin{split}
In[1]:=~~&{\mit A\/}=
      \text{Table}[\,\text{Which}[\,m \text{\,\it ==\,} n \text{\,\it ==\,} 1 \vee  m \text{\,\it ==\,} n \text{\,\it ==\,} 5,
      1,\\
&\phantom{{\mit A\/}=\text{Table}[\text{Which}[\,}
          m \text{\,\it ==\,} 1 \wedge n \text{\,\it ==\,} 2 \vee m \text{\,\it ==\,} 3 \wedge n \text{\,\it ==\,} 4, \tfrac{1}{3}\,,\\
&\phantom{{\mit A\/}=\text{Table}[\text{Which}[\,}
 m \text{\,\it ==\,} 3 \wedge n \text{\,\it ==\,} 2 \vee m \text{\,\it ==\,} 5 \wedge n \text{\,\it ==\,} 4,
 \tfrac{2}{3}\,,\\
 &\phantom{{\mit A\/}=\text{Table}[\text{Which}[\,}
m \text{\,\it ==\,} 2 \wedge n \text{\,\it ==\,} 3,
\tfrac{5}{7}\,, m \text{\,\it ==\,} 4 \wedge n \text{\,\it ==\,}
3, \tfrac{2}{7}\,, \text{True},
          0\,],\\
& \phantom{{\mit A\/}=\text{Table}[}\{m, 5\}, \{n, 5\}\,];\\
 &{\mit B\/} = {\mit A\/} /. \{\tfrac{1}{3}
 \rightarrow \tfrac{5}{7}\,,
 \tfrac{2}{3} \rightarrow
 \tfrac{2}{7}\,,
          \tfrac{5}{7} \rightarrow
          \tfrac{1}{3}\,, \tfrac{2}{7} \rightarrow \tfrac{2}{3}\};\\
  &{\mit C\/} =
      \tfrac{1}{2}\, {\mit A\/} + \tfrac{1}{2} \,{\mit B\/};\\
&{\mit finish\/}[{\mit A\/}\_] :=
    \text{Rationalize}[\,
      \text{N}[\,\text{Nest}[\,(\text{\#} . \text{\#}) \text{\&},  {\mit A\/}, 12\,] . \{0, 0, 1, 0, 0\},
        16\,],\,10^{-20}]\\
 & \{\text{MatrixForm}[{\mit A\/}], \text{MatrixForm}[{\mit B\/}],
    \text{MatrixForm}[{\mit C\/}]\}\\
&\{{\mit finish\/}[{\mit A\/}],{\mit finish\/}[{\mit B\/}], {\mit
finish\/}[{\mit C\/}]\}
\end{split}
\]
\[
\begin{split}
Out[5]=~~&\{
\begin{pmatrix}
  1 & \tfrac{1}{3} & 0 & 0 & 0 \\
  0 & 0 & \tfrac{5}{7} & 0 & 0 \\
  0 & \tfrac{2}{3} & 0 & \tfrac{1}{3} & 0 \\
  0 & 0 & \tfrac{2}{7} & 0 & 0 \\
  0 & 0 & 0 & \tfrac{2}{3} & 1
\end{pmatrix}\,,
\begin{pmatrix}
  1 & \tfrac{5}{7} & 0 & 0 & 0 \\
  0 & 0 & \tfrac{1}{3} & 0 & 0 \\
  0 & \tfrac{2}{7} & 0 & \tfrac{5}{7} & 0 \\
  0 & 0 & \tfrac{2}{3} & 0 & 0 \\
  0 & 0 & 0 & \tfrac{2}{7} & 1
\end{pmatrix}\,,
\begin{pmatrix}
  1 & \tfrac{11}{21} & 0 & 0 & 0 \\
  0 & 0 & \tfrac{11}{21} & 0 & 0 \\
  0 & \tfrac{10}{21} & 0 & \tfrac{11}{21} & 0 \\
  0 & 0 & \tfrac{10}{21} & 0 & 0 \\
  0 & 0 & 0 & \tfrac{10}{21} & 1
\end{pmatrix}
\}\\[1ex]Out[6]=~~ &\{\{\,\tfrac{5}{9}\,,\,0,\,0,\,0,\,\tfrac{4}{9}\,\}\,,
\{\,\tfrac{5}{9}\,,\,0,\,0,\,0,\,\tfrac{4}{9}\,\}\,,
\{\,\tfrac{121}{221}\,,\,0,\,0,\,0,\,\tfrac{100}{221}\,\}\}
\end{split}
\]

%%%%%%%%%%%%%%%%%%%%%%%%%%%%%%
% For BiBTeX users, just uncomment the following two lines
%\bibliographystyle{unsrt}
%\bibliography{mybibfile}

\end{document}